\documentclass{article}
\usepackage[utf8]{inputenc}
\usepackage{geometry}
\usepackage{physics}
\usepackage{dsfont}
\usepackage{amsmath}
\usepackage{amssymb}
\usepackage{graphicx}
\usepackage{xcolor}
\usepackage{wrapfig}
\usepackage{caption}
\usepackage{subcaption}
\usepackage{bbm}
\usepackage{authblk}
\title{Effects of intrinsic decoherence on discord-like correlation measures of two-qubit spin squeezing model}

\date{}
\author[1]{\small Venkat Abhignan}
\author[2]{\small R. Muthuganesan}

\affil[1]{\footnotesize Department of Physics, National Institute of Technology, Tiruchirappalli- 620015, Tamil Nadu, India.}
\affil[2]{\footnotesize Centre for Nonlinear Science and Engineering, School of Electrical and Electronics Engineering, SASTRA Deemed University, Thanjavur-613 401, Tamil Nadu India.}

\begin{document}

\maketitle
\section*{Abstract}
Quantum decoherence happens when the system interacts with the environment. Quantum correlation behaviours in the two-qubit spin squeezing model are studied under the influence of intrinsic decoherence. Quantitative results were determined, which depend on parameters of the physical system by checking different quantifiers of quantum correlation such as entanglement, local quantum uncertainty, trace distance discord and uncertainty-induced quantum nonlocality. We show that the entanglement suffers from intrinsic decoherence and exhibits sudden death, whereas the other measures are more robust against intrinsic decoherence. Further, we highlight the role of spin squeezing coupling constant and magnetic field. 

\section{Introduction}
Nonlocal attributes of a quantum system have been considered as an important resource for various quantum information processing. Only systems in the quantum realm have nonclassical correlations which have no counterparts in the classical regime.  In earlier times, the entanglement was assumed to be the only characteristic trait of such quantum correlations, for it was considered to be the foundational ingredient to violate Bell's inequalities \cite{Schrodinger1935, EPR}. This initial assumption of quantum information led to a wide range of applicability of this concept in quantum computing \cite{NIELSEN}, quantum communications \cite{QT,QT2} and quantum cryptography \cite{QC} to measure nonlocality.  It also fueled the development of quantum technology. However, a paradigm shift happened over the last two decades in measuring quantum correlations \cite{RevModPhys.84.1655} due to the absence of entanglement in one qubit naturally unentangled bipartite systems \cite{ONE_BIT, EXP_WITHOUT_ENTANGLEMENT}. Such experimentally relevant quantum systems, beyond entanglement \cite{nature2012,prl2013} were associated with quantum discord \cite{prl2001,prl2008}. But there exist proliferated ways of describing the presence of quantum discord since it does not have a complete analytical form, which is a hard problem \cite{Huang_2014, qda}. \\ 

To find an appropriate measure and a way to calculate the quantum correlations beyond entanglement for mixed states, geometric quantum discord based on Hilbert–Schmidt norm was introduced \cite{GD2010}. However, due to its contractive nature \cite{contractive} and local ancilla problem \cite{ancilla} modified geometrical quantifier trace distance discord (TDD) is used as a measure of such correlations for two-qubit systems \cite{TDD2014}. It is defined based on the trace norm (Schatten 1-norm), the distance between the state under consideration, and closest zero discord states \cite{tdd2012,tddpra2013}. Also, other prominent quantifiers like local quantum uncertainty (LQU) \cite{Girolami2013} and uncertainty-induced nonlocality (UIN) \cite{UIN2014} are used for the evaluation of entropic uncertainty, which arises in a quantum system due to a local measurement of an observable. These quantifiers are inspired by Wigner–Yanase skew information \cite{Wigner1963,skew2003} where LQU quantifies minimal skew information of a given state and local non-commuting observable and in contrast, UIN quantifies maximal skew information between the given state and local commuting observables.\\ 

In open quantum systems, information is irrevocably lost to the surrounding environments, and physical properties of the system approach a macroscopic scale where quantum behaviours are no longer present. This loss of information is attributed to the decay of quantum coherence as the dynamic system is spread over many degrees of freedom of the interacting environment \cite{prd1981,prd1982,RevModPhys1987} and the decaying dynamics is governed by Lindblad master equation. Another generic way of studying this process is by "intrinsic decoherence", where a simple modification of quantum dynamics leads to phase decoherence without any external reservoir and energy dissipation. Milburn initially proposed a unique way to introduce decoherence in the traditional quantum mechanical system by modification of unitary Schr\"{o}dinger evolution to obtain a Lindblad equation \cite{milburn1991}. Deterioration of coherence through this method has been studied on different physically relevant systems since then. Moya et al. showed that decoherence controls the decay of revivals in the atomic population inversion through the Jaynes-Cummings model of atom-field interaction \cite{Moya}. Zheng and Zhang studied a similar model with Heisenberg exchange interaction to observe the decay of entanglement quantified by the concurrence\cite{Zheng2017}. Mohamed et al. checked the robustness of correlation measures through measurement-induced non-locality (MIN), TDD and Bures entanglement on three-qubit XY chain \cite{MOHAMED2021}. Yang et al. investigated the rate of decay in quantum Fisher information on two-qutrit Heisenberg XY chain \cite{Yang_2017}. Measures based on skew information are in close correspondence to quantum Fisher information, which exploits quantum Chernov bound and produces the statistical uncertainty in parameter estimation \cite{qfi2018}. This relationship shows that LQU and UIN can be ideal tools to quantify correlations as resources in quantum estimation theory and quantum metrology. Recently Muthuganesan and Chandrasekhar studied the dynamics of MIN \cite{Muthuganesan2021}, while Guo et al. investigated dynamics of LQU on two-qubit Heisenberg spin chain described by Dzyaloshinski-Moriya interaction and Heisenberg anisotropic interaction \cite{Guo_2021}. Similarly, Chlih et al. studied the dynamics of LQU on the Heisenberg XXZ spin chain model with Dzyaloshinskii–Moriya interaction. To observe the influence of decoherence parameter in the time evolution process, we considered studying results on quantum discord using the two-qubit spin models as resources \cite{sss1993,sss2003}. Such squeezed spin systems have diverse physical applications like in macroscopic two-state systems of interferometers and Josephson junctions \cite{interferometer1,interferometer2}. Moreover our study is based on one-axis twisting model which is an interesting special case under recent investigation \cite{oat0,oat1,oat2,oat3,oat3.5,oat4,oat5}. This study can lead to interesting insights on the properties of realistic open quantum systems from the perspectives of correlation measures.  \\ 

In this article, we study the dynamics of quantum correlation quantified by entanglement, TDD, LQU and UIN under the intrinsic decoherence. We qualitatively observe similar characteristics as obtained in existing literature using the Milburn scheme of decoherence. Further, it is shown that correlation measures based on the skew information are more robust than entanglement and TDD against intrinsic decoherence. The effects of spin squeezing coupling and external magnetic fields are also studied.
The paper is structured as follows: In Sec. 2, we introduce Milburn's modification of quantum system under intrinsic decoherence with stochastic evolution. Further, we briefly describe the theoretical model, initial state and evolved state of our analysis. In Sec. 3, we define all the quantum correlations measures considered in this article. The quantitative results and discussion are given in Sec. 4. Finally, we present the conclusions in Sec. 5.

\section{Intrinsic decoherence and the spin squeezing model}
 Milburn obtained an evolution by assuming that in sufficiently short time scales, the system undergoes a stochastic sequence of identical unitary transformations \cite{milburn1991}. Accordingly, he derived the first-order correction to the evolution equation for a system with Hamiltonian $H$ and state $\rho(t)$ as \begin{equation}
     \frac{d\rho(t)}{dt} = -i[H,\rho(t)]-\frac{\gamma}{2}\left[H,[H,\rho(t)]\right].
 \end{equation} 
 Here $\gamma$ is the intrinsic decoherence parameter, the mean frequency of unitary steps and becomes the expansion parameter for consecutive corrections at higher orders. If the frequency of the time steps is large enough, the evolution appears continuous on laboratory time scales, and at zeroth order ($\gamma\rightarrow0$) Schr\"{o}dinger equation is recovered. The first-order correction in Eq. (1) with double commutator represents the decoherence effect on the system, induced by diffusion in variables that do not commute with Hamiltonian. We rewrite Eq.(1) in Kraus operator-sum representation using Kraus operators $M_l$ as \begin{equation}
     \rho(t) = \sum_{l=0}^\infty M_l(t)\rho(0)M_l^{\dagger}(t),
 \end{equation}
 where $\rho(0)$ is the initial state of the system and $M_l(t)$ is defined as \begin{equation}
     M_l(t) = \frac{(\gamma t)^{l/2}}{\sqrt{l!}} H^l \exp{-iHt}\exp{- \frac{\gamma t}{2} H^2}.
 \end{equation}
 With this the time evolved state is reformulated as  \begin{equation}
     \rho(t) = \sum_{m,n} \exp{- \frac{\gamma t}{2}(E_m-E_n)^2-i(E_m-E_n)t}\bra{\phi_m}\rho(0)\ket{\phi_n}\ket{\phi_m}\bra{\phi_n},
 \end{equation}
 where $E_{m,n}$ and $\phi_{m,n}$ are the eigenvalues and their corresponding eigenvectors derived from Hamiltonian. Recently this effect of intrinsic decoherence in physical systems was studied, as mentioned earlier \cite{MOHAMED2021,Muthuganesan2021,Guo_2021}. 
 
 To study the quantum correlations and effects of intrinsic decoherence on a physical system we consider the Hamiltonian \begin{equation}
    H = \mu S_x^2+ \zeta S_y^2 + \Gamma(S_x S_y+S_y S_x) + B S_z,
\end{equation}
 which is a solid-state system for two-qubit spin squeezing model \cite{sss1993,sss2003} where $\mu\geq0$, $\zeta\geq0$ and $\Gamma\geq0$ are the strength of the spin squeezing interactions in $x$, $y$ and $x-y$ directions with external magnetic field $B(\geq0)$ in $z$-direction. Also here $S_\alpha=\frac{1}{2}\sum_{i=1}^2\sigma_\alpha^i(\alpha=x,y,z)$ are collective spin operators where $\sigma_\alpha^i$ are the Pauli matrices of $i$th spin. If $\zeta=\Gamma=0$, the Hamiltonian reduces to one axis twisting model with a tranverse field such as \begin{equation}
    H = \mu S_x^2 + B S_z.
\end{equation}
As mentioned, this considered system recently gained interest in quantum information processing \cite{oat1,oat2}. Solving the Schr\"{o}dinger equation $H\ket{\phi}=E\ket{\phi}$, we obtain the eigenvalues of the above Hamiltonian and their corresponding eigenvectors in the standard basis $\{\ket{00},\ket{01},\ket{10},\ket{11}\}$ as: \begin{subequations}
\begin{align}
    E_1=  \frac{\mu-\kappa}{2}, ~& ~~~~~~\ket{\phi_1}= \frac{\left((2B-\kappa)\ket{11}+\mu\ket{00}\right)}{\sqrt{\mu^2+(2B-\kappa^2)}},~~~~~~~~~~~~\\E_2=  \frac{\mu+\kappa}{2}, ~& ~~~~~~\ket{\phi_2}= \frac{\left((2B+\kappa)\ket{11}+\mu\ket{00}\right)}{\sqrt{\mu^2+(2B+\kappa^2)}},~~~~~~~~~~~~\\E_3=  0, ~& ~~~~~~\ket{\phi_3}= \frac{\left(\ket{10}-\ket{01}\right)}{\sqrt{2}},~~~~~~~~~~~~\\E_4= \mu, ~& ~~~~~~\ket{\phi_4}= \frac{\left(\ket{10}+\ket{01}\right)}{\sqrt{2}},~~~~~~~~~~~~
\end{align}
\end{subequations}
where $\kappa=\sqrt{\mu^2+4B^2}$. To study the dynamical behaviours of quantum correlation under the influence of intrinsic decoherence (Such as in Eq.(1)), we consider the well-known X-state as an initial state for our analysis in the basis $\{\ket{00},\ket{01},\ket{10},\ket{11}\}$. Using Eqs. (7) and (4) we compute the time evolved density matrix in the same two-qubit computational basis as 
\begin{equation}
    \rho(t) = \left(\begin{array}{cccc}
\rho_{11}(t) & 0 & 0 & \rho_{14}(t) \\
0 & \rho_{22}(t) & \rho_{23}(t) & 0\\
0 & \rho_{32}(t) & \rho_{33}(t) & 0\\
\rho_{41}(t)  & 0 & 0 & \rho_{44}(t) 
\end{array}\right).
\end{equation}
  
 \section{Quantum correlation measures}
 In this section, we review the quantum correlation measures to be studied in this article. To quantify the quantum correlation contained in a bipartite system, we consider a state $\rho$ shared between the subsystems $\rho^a$ and $\rho^b$ in the separable Hilbert space $\mathcal{H}^{ab}=\mathcal{H}^{a} \otimes \mathcal{H}^{b}$.
\subsection{Concurrence}
The concurrence quantifies the degree of entanglement in the considered state $\rho$ and is defined as \cite{hill1997}
\begin{equation}
    C(\rho) = \max \{0, \lambda_1 - \lambda_2 - \lambda_3 - \lambda_4\}, 
\end{equation}
where $\lambda_i$ are the eigenvalues of matrix $R=\sqrt{\sqrt{\rho}\tilde{\rho}\sqrt{\rho} }$ and arranged in descending numerical values. Further, the spin flipped matrix is $\tilde{\rho}=(\sigma_y \otimes \sigma_y) \rho^* (\sigma_y \otimes \sigma_y)$ and * denotes the complex conjugate in computational basis. 
The concurrence of the time evolved state $\rho(t)$ in Eq. (8) can be calculated from evaluated expression \begin{equation}
    C(\rho(t)) = 2 \max \{0,|\rho_{14}(t)|-\sqrt{(\rho_{33}(t)\rho_{22}(t))},|\rho_{23}(t)|-\sqrt{(\rho_{44}(t)\rho_{11}(t))}\}.
\end{equation}
It is worth mentioning that the $C(\rho)$ varies from 0 to 1; the minimal and maximal values correspond to unentangled and maximally entangled states, respectively.
\subsection{Local Quantum Uncertainty }
Next, we employ LQU as the second quantifier of bipartite quantum correlation. LQU is one of the reliable quantifiers of minimal quantum uncertainty of the state $\rho$ due to the measurement of a local observable \cite{Girolami2013}. It is defined as 
\begin{align}
\mathcal{U}(\rho) =-~^{\text{min}}_{K ^{a}}~\text{Tr}[\sqrt{\rho},K^a\otimes I]^2
\end{align}
where the minimization is taken over the local observable $K^a$. The analytical evaluation of LQU to measure the amount of quantum correlations in the state $\rho(t)$ requires Fano-Bloch decomposition \cite{Habiballah2018} \begin{equation}
    \rho = \frac{1}{4} \sum_{\alpha,\beta} R_{\alpha \beta} \sigma_\alpha \otimes \sigma_\beta.
\end{equation}
Similarly, we can obtain the Fano-Bloch decomposition of matrix $\sqrt{\rho(t)}$ with $\mathcal{R}_{\alpha \beta} = \Tr(\sqrt{\rho(t)} \sigma_\alpha \otimes \sigma_\beta)$ as \begin{equation*}
    \sqrt{\rho} = \frac{1}{4} \sum_{\alpha,\beta} \mathcal{R}_{\alpha \beta} \sigma_\alpha \otimes \sigma_\beta
\end{equation*}
The analytical formula of LQU is 
\begin{equation}
    \mathcal{U}(\rho) = 1 - \text{max} \{\Lambda_1, \Lambda_2, \Lambda_3\},
\end{equation}
where $\Lambda_i$ are the eigenvalues of matrix W whose matrix elements are given by $w_{ij} \equiv \Tr\{\sqrt{\rho}(\sigma_i\otimes\mathbb{I}_2)\sqrt{\rho}(\sigma_i\otimes\mathbb{I}_2)\}$, with $i,j=1,2,3$. 
 With this, we can determine matrix W where the diagonal elements and off-diagonal elements are given by
    \begin{equation*}
        w_{ii} = \frac{1}{4} \left[\sum_{\beta}\left(\mathcal{R}^2_{0\beta}-\sum_{k}\mathcal{R}^2_{k\beta}\right)\right]+\frac{1}{2}\sum_{\beta}\mathcal{R}^2_{i\beta}\,\,\,\,\,\, \hbox{and}\,\,\,\,\,\, w_{i j} = \frac{1}{2}\mathcal{R}_{i\beta} \mathcal{R}_{j\beta}\,\,\,\,\,(i \neq j), \hbox{respectively}.
    \end{equation*}
    Consecutively we can obtain the eigenvalues $\Lambda_i$ of matrix W as \begin{equation}
        \Lambda_1 = \frac{1}{2}(w_{11}+w_{22})+\sqrt{w_{12}^2+\frac{1}{4}(w_{11}-w_{22})^2}, \Lambda_2 = \frac{1}{2}(w_{11}+w_{22})-\sqrt{w_{12}^2+\frac{1}{4}(w_{11}-w_{22})^2}\,\,\,\,\, \hbox{and}\,\,\,\,\, \Lambda_2 = w_{33}.
    \end{equation} 
    Consequently, using this, we obtain LQU from Eq. (13) for the state $\rho(t)$.
   \subsection{Trace distance discord}
    The next correlation measure is Trace distance discord (TDD), it is defined as trace norm distance between state $\rho$  and the set of classical-quantum density matrices, which exhibit zero quantum discord with respect to local measurements on subsystem $a$ \cite{TDD2014}. It is defined as 
    \begin{equation}
\mathcal{T}(\rho ) =~^{\text{min}}_{\Pi ^{a}}\| \rho - \Pi ^{a}(\rho )\| . 
\end{equation}
where the minimization taken over local measurements on subsystem $a$.

    TDD is also found to be more sensitive and stronger measure compared to LQU and concurrence \cite{Khedif_2019}. Analytically closed form of TDD $\mathcal{T}$ is solved in Fano-Bloch components in Eq. (12) as  \begin{equation}
      \mathcal{T}(\rho) = \sqrt{\frac{R_{11}^2 R_{max}^2 - R_{22}^2 R_{min}^2}{R_{max}^2 - R_{min}^2 + R_{11}^2 - R_{22}^2}}.
  \end{equation}
  In terms of the X-state elements of $\rho(t)$ it is written in compact form as
\begin{equation}
    \mathcal{T}(\rho)=\frac{1}{2}\sqrt{\frac{\gamma_1^2 \max\{\gamma_3^2,\gamma_2^2+x_{A3}^2\}-\gamma_2^2 \min\{\gamma_3^2,\gamma_1^2\}}{\max\{\gamma_3^2,\gamma_2^2+x_{A}^2\}- \min\{\gamma_3^2,\gamma_1^2\}+\gamma_1^2-\gamma_2^2}}
\end{equation}
where \begin{equation}
    \gamma_1 = 2(\rho_{32}+\rho_{41}),\,\gamma_2 = 2(\rho_{32}-\rho_{41}),\,\gamma_3 = 1-2(\rho_{22}+\rho_{33}) \,\,\,\hbox{and}\,\,\, x_A = 2(\rho_{11}+\rho_{22})-1.
\end{equation}
 \subsection{Uncertainty-induced quantum nonlocality}
    UIN is another quantifier which is enhanced MIN \cite{MIN2011} and shares similar characteristics with LQU \cite{UIN2014}. UIN quantifies the maximal skew information between the state $\rho$ and local commuting observable $K^a$ which acts on subsystem $a$. It is defined as \begin{equation}
       \mathcal{U_C}(\rho) = ~^{\text{max}}_{K ^{a}} \mathcal{I}(\rho, K^a).
    \end{equation} Here maximum is taken over all the local observables $K^a=K^a_A \otimes \mathbbm{1}_b $ where $K^a_A$ refers to an Hermitian operator on $a$ and $\mathbbm{1}_b$ is $2 \times 2$ identity operator corresponding to subsystem $b$. $K^a$ is taken such as it commutes with reduced density matrix $\rho_a$ of $a$. In some cases it is shown to reveal the nonclassical correlations even when entanglement and LQU are absent \cite{Khedif2021}. The closed form for UIN for the state $\rho(t)$ is derived from matrix W and its eigenvalues $\Lambda_i$ in Eq. (14) used for computing LQU. The expression for UIN is 
        \begin{equation}
\mathcal{U_C}(\rho(t)) = \Bigg\{\begin{array}{l}
   \ 1 - \min\{\Lambda_1, \Lambda_2, \Lambda_3\} \ \ ,\ \ \vec{\mathbf{r'}} = \vec{\mathbf{0}} \\ 
  \ 1 - \frac{1}{|\vec{\mathbf{r'}}|^2} \vec{\mathbf{r'}} W \vec{\mathbf{r'}}^T \ \ ,\ \ \vec{\mathbf{r'}} \neq \vec{\mathbf{0}}
  \end{array}
  \end{equation}
  where $\vec{\mathbf{r'}}^T$ is the transpose of Bloch vector $\vec{\mathbf{r'}}$ with norm $|\vec{\mathbf{r'}}|$.
 \section{Results and Discussion}
To check the effect of intrinsic decoherence on different quantum correlations, we make a comparative study by varying system parameters, spin squeezing interaction $\mu$, external magnetic field $B$  and decoherence parameter $\gamma$. We compute correlations from the analytical expressions and plot the Concurrence, LQU, UIN and TDD measures for varying time $t$. As mentioned in the plots, we have scaled the correlation measures by a suitable numerical factor for easy comparison and better visibility. \\
First, we consider the Bell diagonal state with maximally mixed marginal which is defined as  
\begin{align}
    \rho(0) = \frac{1}{4} \left( \mathds{1}_a\otimes\mathds{1}_b+\sum^3_{j=1}c_j\sigma_a^j \otimes \sigma_b^j\right)
\end{align}
where $\sigma^j_i$ are Pauli spin matrices and  $\vec{c}=(c_1,c_2,c_3) $ is a three dimensional vector composed of correlation coefficients such that $-1\leq c_j \leq 1$ completely specifies the quantum state. Initial investigation is on this X-structured averaged maximally entangled state as in Eq. (8) and non-zero elements of time evolved density matrix are

\begin{equation}
  \begin{gathered}
    \rho_{11}(t) = A - \frac{2\mu BC}{\kappa^2}\left[1-cos(\kappa t)\exp{-\frac{\gamma t}{2}\kappa^2}\right], \\ \rho_{44}(t) = A + \frac{2\mu BC}{\kappa^2}\left[1-cos(\kappa t)\exp{-\frac{\gamma t}{2}\kappa^2}\right],
  \\ \rho_{14}(t) = \rho^*_{41}(t) = \frac{\mu^2C}{\kappa^2} + \frac{2BC}{\kappa^2}\exp{\frac{-\gamma t \kappa^2}{2}}\left[2B cos(\kappa t)+i\, \kappa\, sin(\kappa t)\right],  \\ 
  \rho_{22}(t) = \rho_{33}(t) = D , \\
  \rho_{23}(t) = \rho_{32}(t) = E ,
  \end{gathered}
\end{equation}
where $A=\frac{1}{4}(1+c_3)$, $C=\frac{1}{4}(c_1-c_2)$, $D=\frac{1}{4}(1-c_3)$, $E=\frac{1}{4}(c_1+c_2)$.

The entanglement and quantum correlations for the above time evolved states are zero for the vector $\vec{c}=(0, 0, 0) $ at which the state $\rho(0)=\mathds{1}/4$ is the maximally mixed state.

To understand the dynamical behaviours of quantum correlation, we consider a mixed state vector $\vec{c}=(0.9, -0.4, 0.4) $. From Fig. 1, it can be observed that the dynamic quantum correlations have a decreasing procession with an oscillating behaviour and the decay due to the exponential damping factor ($\exp{-\gamma t\kappa^2/2}$). They reach a particular steady-state value after a finite time. In other words, the oscillating correlations of the bipartite state reach the correlating steady state. This steady-state correlation has practical applications in quantum information processing. While comparing the measures, we observe that TDD is pronounced more qualitatively. The same trend is seen in Fig. 2 and Fig. 3. From comparing Fig. 1(a) and Fig. 1(b), the wiggling nature of correlations decreases sooner with increasing the value of the magnetic field and finite steady-state value is attained at an earlier time. Also, usually in the presence of a magnetic field, the correlation strength decreases, but the magnetic field unconventionally strengthens the correlations. In the asymptotic limit, the correlation measures are a function of the ratio $\mu/B$, and the intrinsic decoherence parameter does not affect the steady-state.

To examine the role of the spin squeezing parameter, we have plotted the correlation measures in Fig. 2 for different values of $\mu$. It is observed that the strength of correlations decreases with increasing the value of $\mu$ and reaches the steady-state early for higher values of $\mu$. Here also, we observe the decaying behaviours of quantum correlation due to the damping factor.

Further, from Fig. 3 varying the value of $\gamma$, it is deduced asymptotic steady state is independent of $\gamma$. Increasing the value of $\gamma$ causes the correlations to wiggle less and reach a steady-state sooner. In other words, the decoherence parameter accelerates the correlation measure to achieve a steady-state value. The relevant steady-state correlation measures are obtained, which are only dependent on the parameters of the system. Similar observations were recorded for other correlation measures in Heisenberg spin systems \cite{Guo_2021}.
  \begin{figure}[!ht]
\centering
\begin{subfigure}{0.497\textwidth}
\includegraphics[width=1\linewidth, height=5cm]{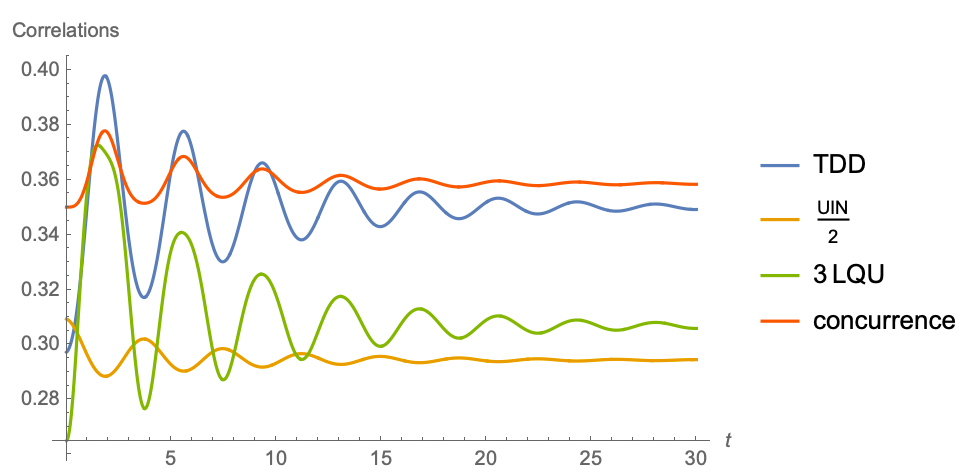} 
\caption{Comparing correlations for $B=0.25$}

\end{subfigure}
\begin{subfigure}{0.497\textwidth}
\includegraphics[width=1\linewidth, height=5cm]{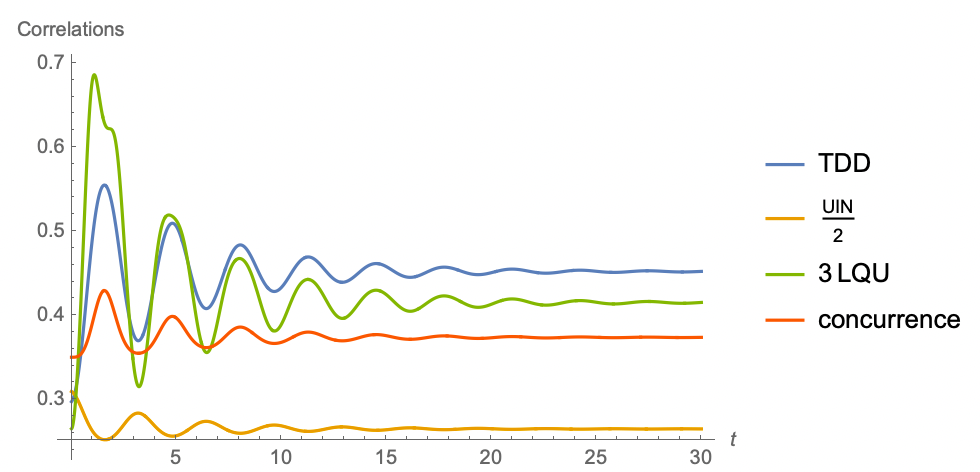}
\caption{Comparing correlations for $B=0.55$}

\end{subfigure}

\caption{Illustrating the dynamic behaviour of TDD, UIN, LQU and concurrence for $\mu=1.6$, $\gamma=0.1$, $c_1=0.9$, $c_2=-c_3=0.4$.}

\end{figure}
\begin{figure}[!ht]
\centering
\begin{subfigure}{0.497\textwidth}
\includegraphics[width=1\linewidth, height=5cm]{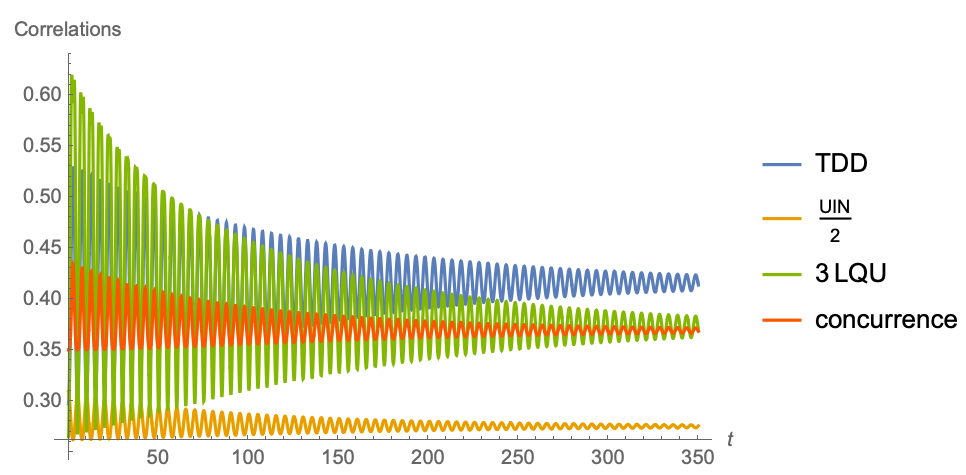} 
\caption{Comparing correlations for $\mu=1.1$}

\end{subfigure}
\begin{subfigure}{0.497\textwidth}
\includegraphics[width=1\linewidth, height=5cm]{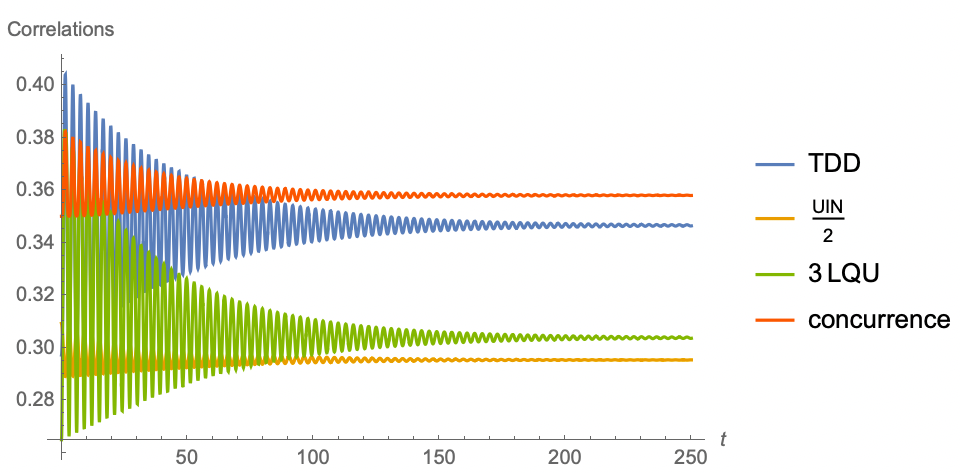}
\caption{Comparing correlations for $\mu=2$}

\end{subfigure}

\caption{Illustrating the dynamic behaviour of TDD, UIN, LQU and concurrence for $B=0.3$, $\gamma=0.01$, $c_1=0.9$, $c_2=-c_3=0.4$.}

\end{figure}
\begin{figure}[!ht]
\centering
\begin{subfigure}{0.497\textwidth}
\includegraphics[width=1\linewidth, height=5cm]{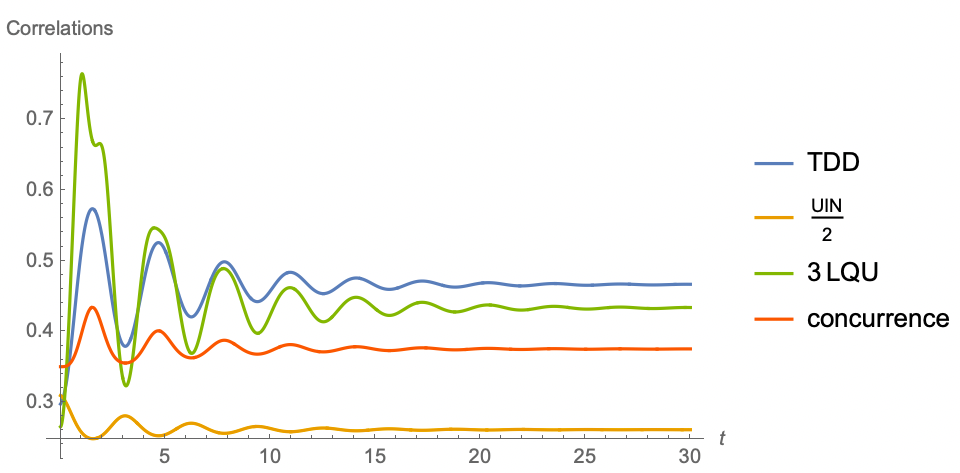} 
\caption{Comparing correlations for $\gamma=0.1$}

\end{subfigure}
\begin{subfigure}{0.497\textwidth}
\includegraphics[width=1\linewidth, height=5cm]{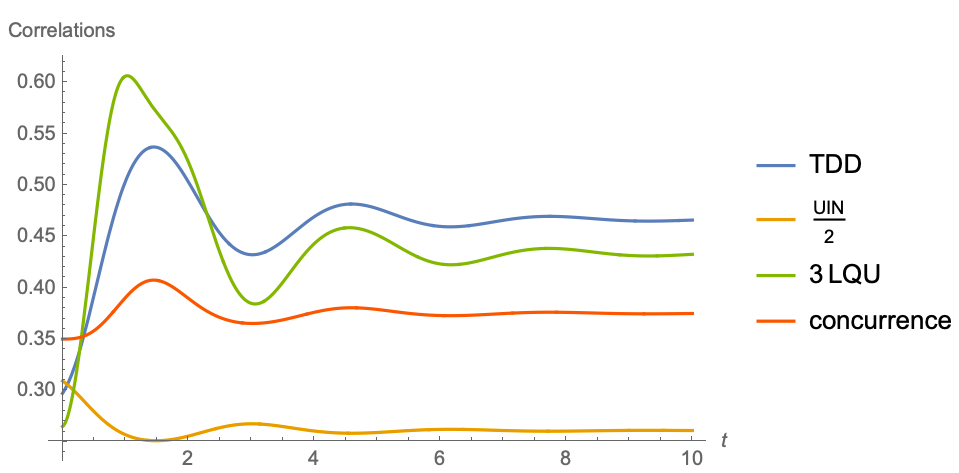}
\caption{Comparing correlations for $\gamma=0.25$}

\end{subfigure}

\caption{Illustrating the dynamic behaviour of TDD, UIN, LQU and concurrence for $\mu=1.6$, $B=0.6$, $c_1=0.9$, $c_2=-c_3=0.4$.}

\end{figure}
 
  Secondly we consider two-qubit Werner state  as
\begin{equation}
      \rho(0) = \frac{1}{4}(1-r)I_{4\times4}+r\left|\varphi\right>\left<\varphi\right|.
\end{equation}
where $\left|\varphi\right> = \frac{1}{\sqrt{2}}\left(\left|00 \right>+\left|11 \right>\right)$ and $r$ characterizes the purity of the state. If $r=1$, the state is a pure state and for $r=0$, state is maximally mixed state. Under the intrinsic decoherence, the matrix elements of the time evolved state are 
\begin{equation}
  \begin{gathered}
    \rho_{11}(t) = b - \frac{2\mu B d}{\kappa^2}\left[1-cos(\kappa t)\exp{-\frac{\gamma t}{2}\kappa^2}\right], \\ \rho_{44}(t) = b + \frac{2\mu B d}{\kappa^2}\left[1-cos(\kappa t)\exp{-\frac{\gamma t}{2}\kappa^2}\right],
  \\ \rho_{14}(t) = \rho^*_{41}(t) = \frac{\mu^2d}{\kappa^2} + \frac{2Bd}{\kappa^2}\exp{\frac{-\gamma t \kappa^2}{2}}\left[2B cos(\kappa t)-i\, \kappa\, sin(\kappa t)\right],  \\
  \rho_{22}(t) = \rho_{33}(t) = a, 
  \end{gathered}
\end{equation}
where $a=\frac{1}{4}(1-r)$,$b=\frac{1}{4}(1+r)$ and $d=\frac{r}{2}$.

Compared to the earlier case, one can observe similar dynamic behaviour of correlation measures. The characteristics of quantum correlation depend on the purity of the state $r$. To show this vividly, we set  $r=0.9$ and $\gamma=0.01$, plotting all the measures in Fig. 4. We observe that the measures are oscillating with time and are decreasing exponentially due to the damping factor. After a finite time, the correlation measures reach a steady state. If we increase the strength of the decoherence parameter, quantum correlation measures reach the steady-state in early time compared to lower values of $\gamma$. Further, the steady quantum state is independent of $\gamma$ and functions of $\mu/B$.

To assert the role of $r$, we study the quantum correlation measure for different values of $r$ as a function of $\mu$ and $B$ in Figs. 5 and 6. For $r=0.5$, we observe that the correlation measures show similar oscillating and decaying behaviour. Contrary to the LQU and UIN, entanglement exhibits the well-known behaviour, namely called sudden death of entanglement and intrinsic decoherence, which cause sudden death and birth in TDD. On the other hand, the measures UIN and LQU are more pronounced than the entanglement and TDD.

As observed, the concurrence reaches steady-state slowly for $r=0.9$ in Figs. 4(a), 5(a), 6 compared to $r=0.5$ in Figs. 4(b), 5(b). In which we can detect the presence of other correlation measures even when concurrence is absent. We infer for more pure states that concurrence and TDD have stronger correlation strength in this case and take a longer time to attain correlating steady-state while LQU and UIN attain steady state sooner. 

\section{Conclusion}
We have studied the dynamics of bipartite quantum correlations quantified by the entanglement, LQU, TDD and UIN in the spin squeezing model under the influence of intrinsic decoherence. For two unique initial states, we find that the intrinsic decoherence parameter does not influence the final steady-state, while it does influence the decoherence speed to the steady-state. This quantitative study using different measures provides effective tools for probing models under intrinsic decoherence. We study varying state and system parameters to observe that quantifiers based on Wigner-Yanase skew information are more useful than entanglement and measure based on trace norm in this case. In other words, skew information correlation measures are more robust against intrinsic decoherence compared to other measures.  

 Further, our investigations underscore that the quantum information processing such as metrology and parameter estimation based on the skew information correlations offer more resistance to the effect of intrinsic
decoherence. 

\begin{figure}[!ht]
\centering
\begin{subfigure}{0.497\textwidth}
\includegraphics[width=1\linewidth, height=5cm]{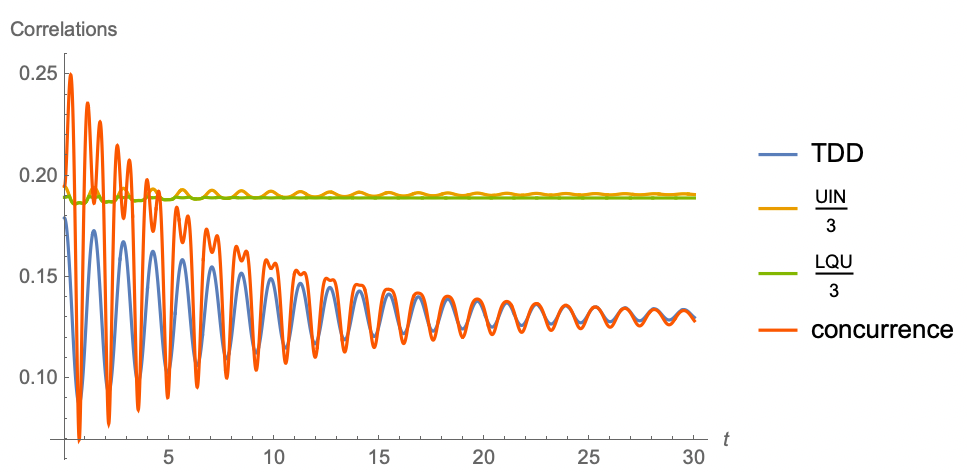} 
\caption{Comparing correlations for decoherence parameter $\gamma=0.01$}

\end{subfigure}
\begin{subfigure}{0.497\textwidth}
\includegraphics[width=1\linewidth, height=5cm]{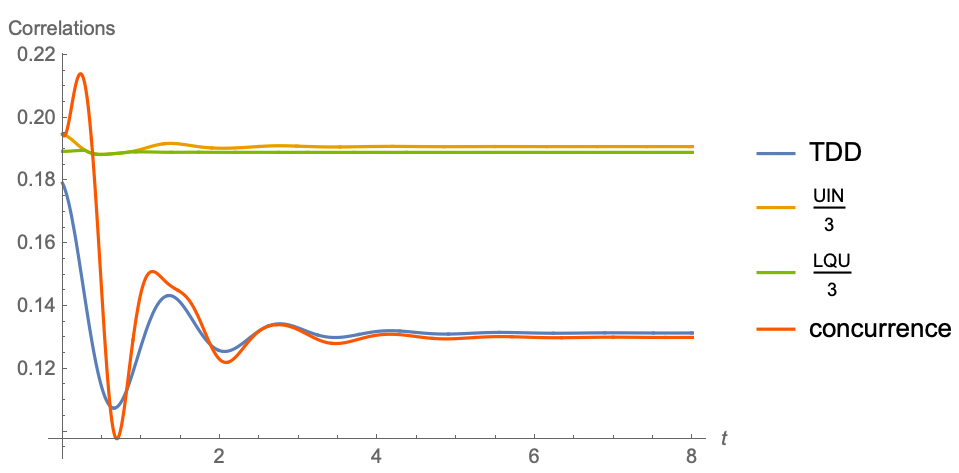}
\caption{Comparing correlations for decoherence parameter $\gamma=0.1$}

\end{subfigure}

\caption{Illustrating the dynamic behaviour of TDD, UIN, LQU and concurrence for $B=0.6$, $\mu=2$ in case 2, $r=0.9$.}

\end{figure}

  \begin{figure}[!ht]
\centering
\begin{subfigure}{0.497\textwidth}
\includegraphics[width=1\linewidth, height=5cm]{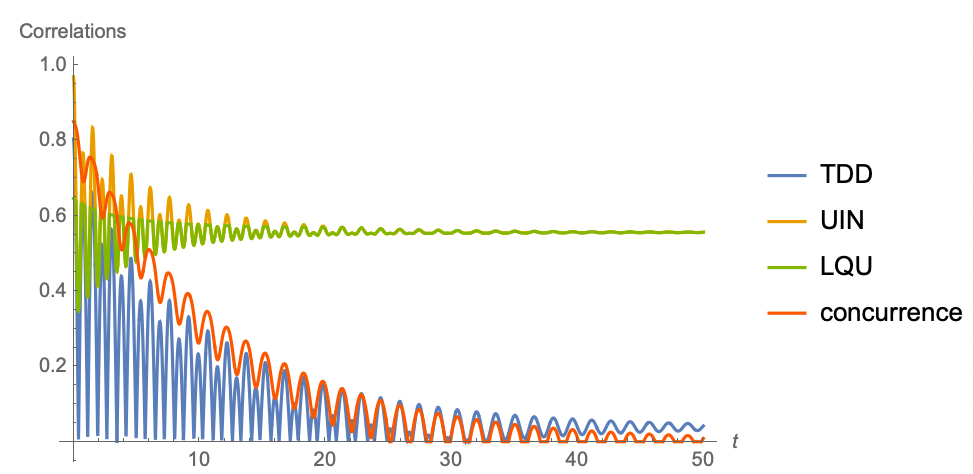} 
\caption{Comparing correlations for case 2, $r=0.9$}

\end{subfigure}
\begin{subfigure}{0.497\textwidth}
\includegraphics[width=1\linewidth, height=5cm]{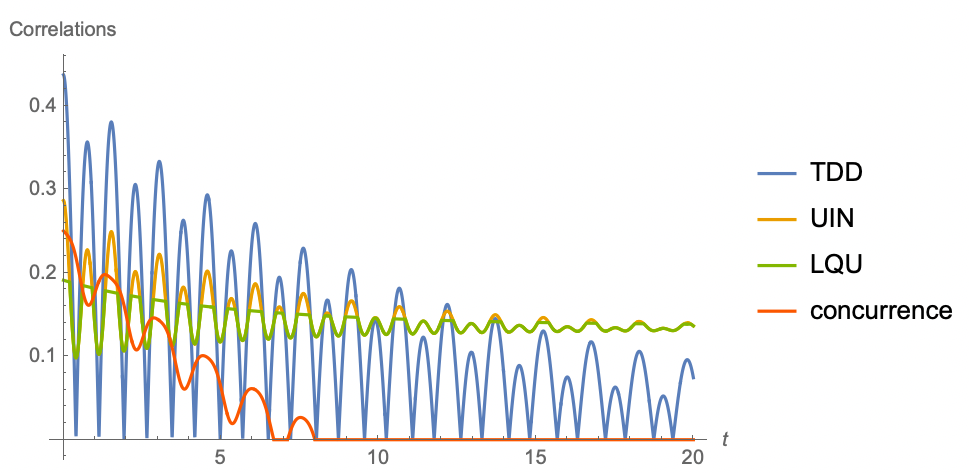}
\caption{Comparing correlations for case 2, $r=0.5$}

\end{subfigure}

\caption{Illustrating the dynamic behaviour of TDD, UIN, LQU and concurrence for $B=2$, $\gamma=0.01$, $\mu=1$.}

\end{figure}
\begin{figure}[!ht]
\centering
\begin{subfigure}{0.497\textwidth}
\includegraphics[width=1\linewidth, height=5cm]{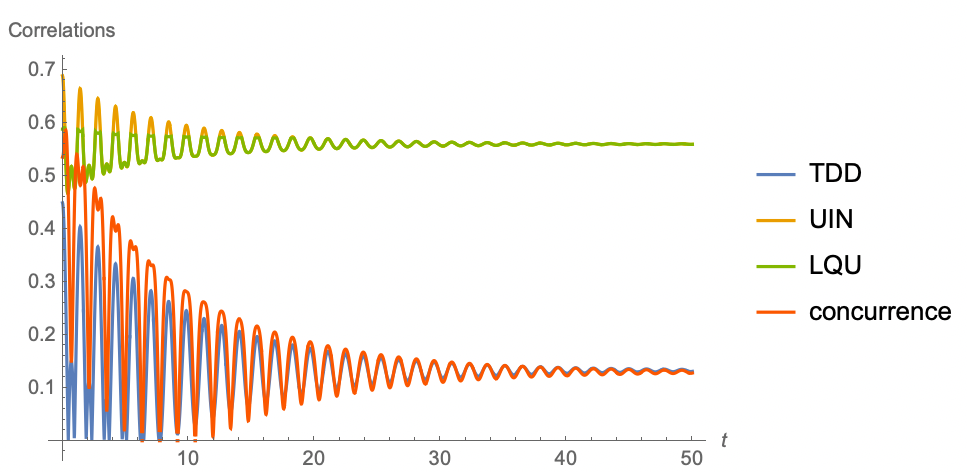} 
\caption{Comparing correlations for case 2, $r=0.9$}

\end{subfigure}
\begin{subfigure}{0.497\textwidth}
\includegraphics[width=1\linewidth, height=5cm]{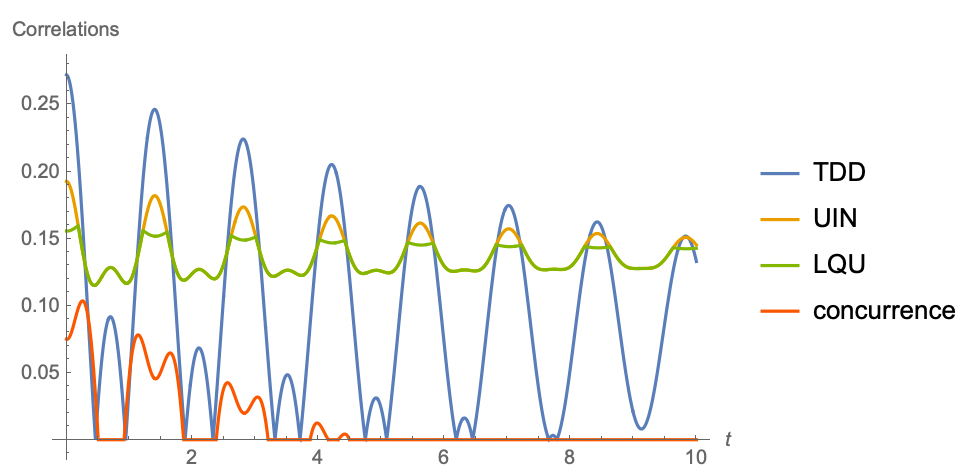}
\caption{Comparing correlations for case 2, $r=0.5$}

\end{subfigure}

\caption{Illustrating the dynamic behaviour of TDD, UIN, LQU and concurrence for $B=1.5$, $\gamma=0.01$, $\mu=2$.}

\end{figure}

\bibliographystyle{ieeetr}
\bibliography{sample.bib}
\end{document}